\documentclass{ptapap}
\usepackage{amsmath}
\usepackage{natbib}

\author{Robert Jaros}[UMK]
\author{Miljenko \v{C}emelji\'{c}}[CAMK,TAI]
\author{W\l odek Klu\'{z}niak}[CAMK]
\author{Dejan Vinkovi\'{c}}[SSI]
\author{Cezary Turski}[UW]
\affil[CAMK]{Nicolaus Copernicus Astronomical Center, Polish Academy of Sciences, Bartycka 18, 00--716
Warsaw, Poland}
\affil[UMK]{Nicolaus Copernicus University, Grudziadzka 5, 87-100 Toru\'{n}, Poland}
\affil[TAI]{Institute of Astronomy and Astrophysics, Academia Sinica, P.O.  Box 23--141, Taipei 106, Taiwan}
\affil[UW]{Faculty of Physics, University of Warsaw, Pasteura 5, 02--093, Warsaw, Poland}
\affil[SSI]{Science and Society Synergy Institute, Bana Josipa Jela\v{c}i\'{c}a 22,HR-40000, \v{C}akovec, Croatia}

\title{Transport of dust grain particles in the accretion disk}

\begin{document}

\maketitle

\begin{abstract}

Entrainment of dust particles in the flow inside and outside of the
proto-planetary disk has implications for the disk evolution and composition
of planets.  Using quasi-stationary solutions in our star-disk simulations
as a background, we add dust particles of different radii in post-processing
of the results, using our Python tool DUSTER.  The distribution and motion
of particles in the disk is followed in the cases with and without the
backflow in the disk.  We also compare the results with and without the
radiation pressure included in the computation.

\end{abstract}

\section{Introduction}

Entrainment of the dust particles in the flow inside and outside of the
proto-planetary disk influences the disk evolution and composition of
planets.  With our newly developed Python tool DUSTER, we performed
post-processing of the results from the PLUTO code \citep{m07,m12}
simulations obtained in \cite{cem19}, by adding dust
particles of different radii. Similar results were obtained using the
expressions for the disk solutions from \cite{cem18},
instead of results from the simulations.  Any other disk and corona model
can easily be supplied as the DUSTER input.

We perform computations with
DUSTER in the cases with and without backflow in the disk. Such backflow
appears in the analytical HD solutions in \cite{KK00}, and is visible
also in the numerical simulations. Trajectories of
four kinds of particles are followed, with the radii $a$ of 0.1, 0.5, 1 and
2 $\mu m$.  We compute the paths of 25 particles of each kind.  Since the
particles melt on approach to the star closer than some critical distance
\citep{vink06,vink09,vink12}, we add the lost particles at the disk
outer rim, so that there are always 25 particles inside the computational
box.  The paths of the melted particles are erased.  Only the particles that
remain in the disk or are pushed away from the star out of the computational
domain are followed.  The trajectories of the particles are computed with
the equation for the acceleration of particles, with gravity, gas drag and
radiation pressure:\\

\begin{equation}
\ddot{\vec{r}}=-G\frac{M_\star}{r^3}\vec{r}-
\frac{\rho_{\mathrm gas}}{\rho_{\mathrm gr}}\frac{c_{\mathrm s}}{a}
(\dot{\vec{r}}-\vec{v}_{\mathrm gas})+\vec{\beta} G\frac{M_\star}{r^2}
\end{equation}\\

The coefficients are \citep{vink06}:\\

\begin{equation}
\beta=0.4\frac{L_\star}{L_\odot}\frac{M_\odot}{M_\star}
\frac{3000\frac{\mathrm kg}{\mathrm m^3}}{\rho_{\mathrm gr}}\frac{\mu m}{a},\
R_{\mathrm in}=0.0344\Psi\left(\frac{1500K}{T_{\mathrm gr}}\right)^2
\sqrt{\frac{L_{\mathrm tot}}{L_\odot}}{\mathrm{[AU]}},
\end{equation}

with the correction factor for the diffuse heating from the dust itself,
$\Psi\sim2$. To test the influence of radiation pressure on the distribution of
particles, we perform computations with and without the radiation
pressure--see Fig.  3

\section{Results}

We show examples with trajectories of particles, computed in cases with and
without backflow in the disk.  We also show an example with computation
performed without the radiation pressure.
\begin{figure}
    \centering \includegraphics[width=0.7\columnwidth ]{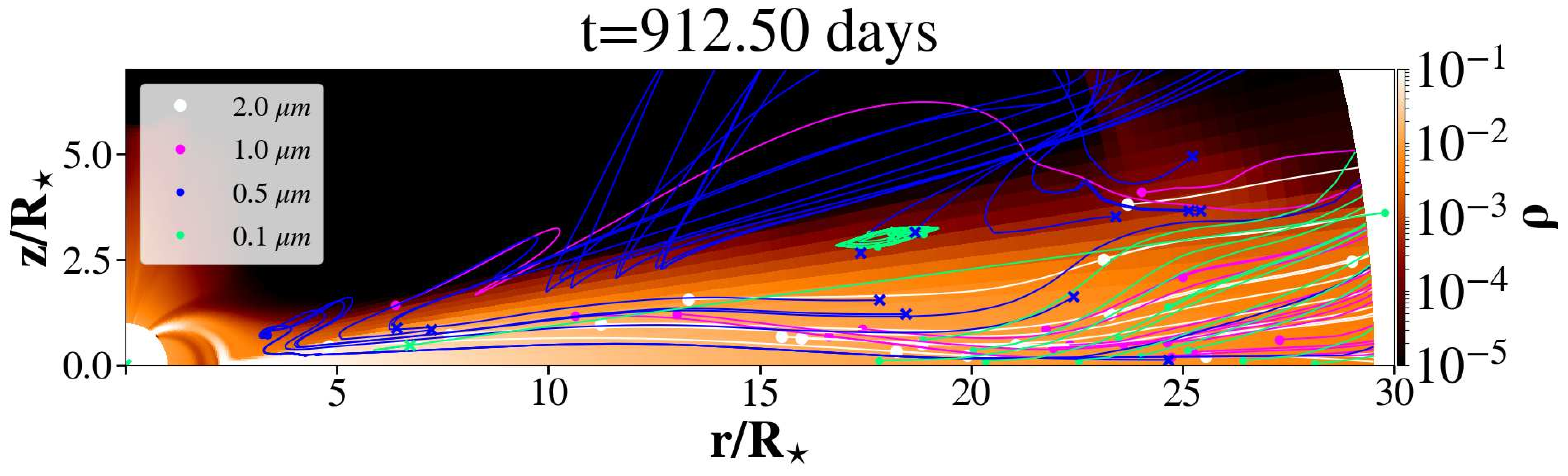}
    \includegraphics[width=0.7\columnwidth]{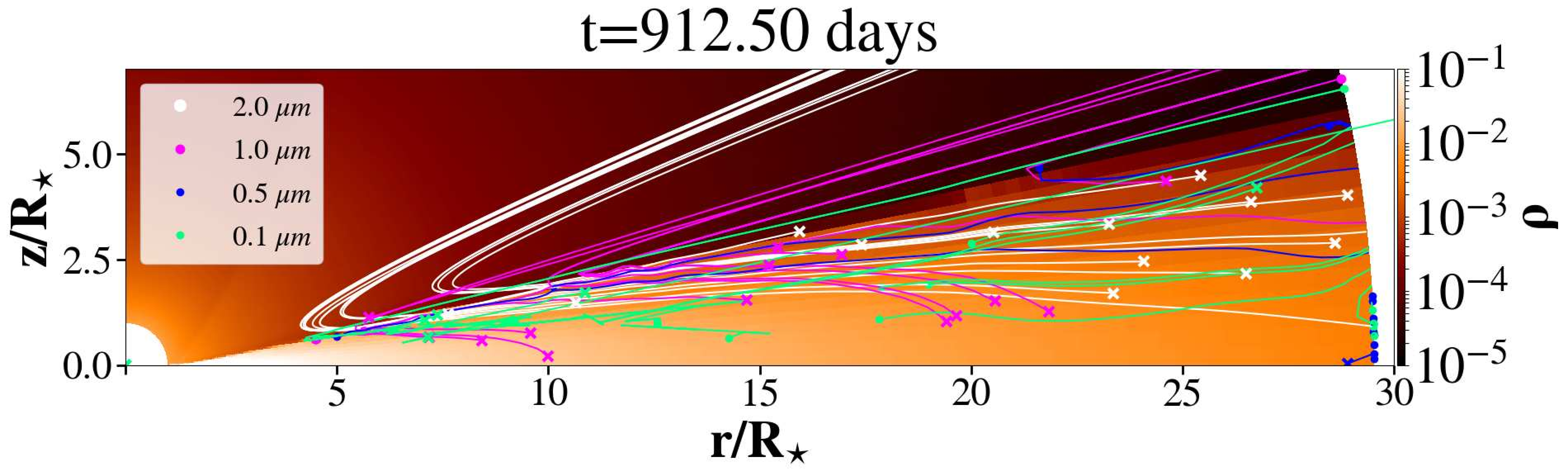} \caption{\small{Particles
    and their trajectories in a case without backflow in the disk (top
    panel) and with backflow in the disk (bottom panel).  Radiation pressure
    is included in both cases, with a fully transparent disk.  Paths of
    particles that melted after approaching the star to the critical
    distance of 4.5 stellar radii were erased, shown are only particles
    pushed away from the star, or those which remained inside the disk.}}
    \label{fig:my_label}
\end{figure}

\begin{figure}
  \centering
  \begin{minipage}{0.48\textwidth}
    \includegraphics[width = 6cm]{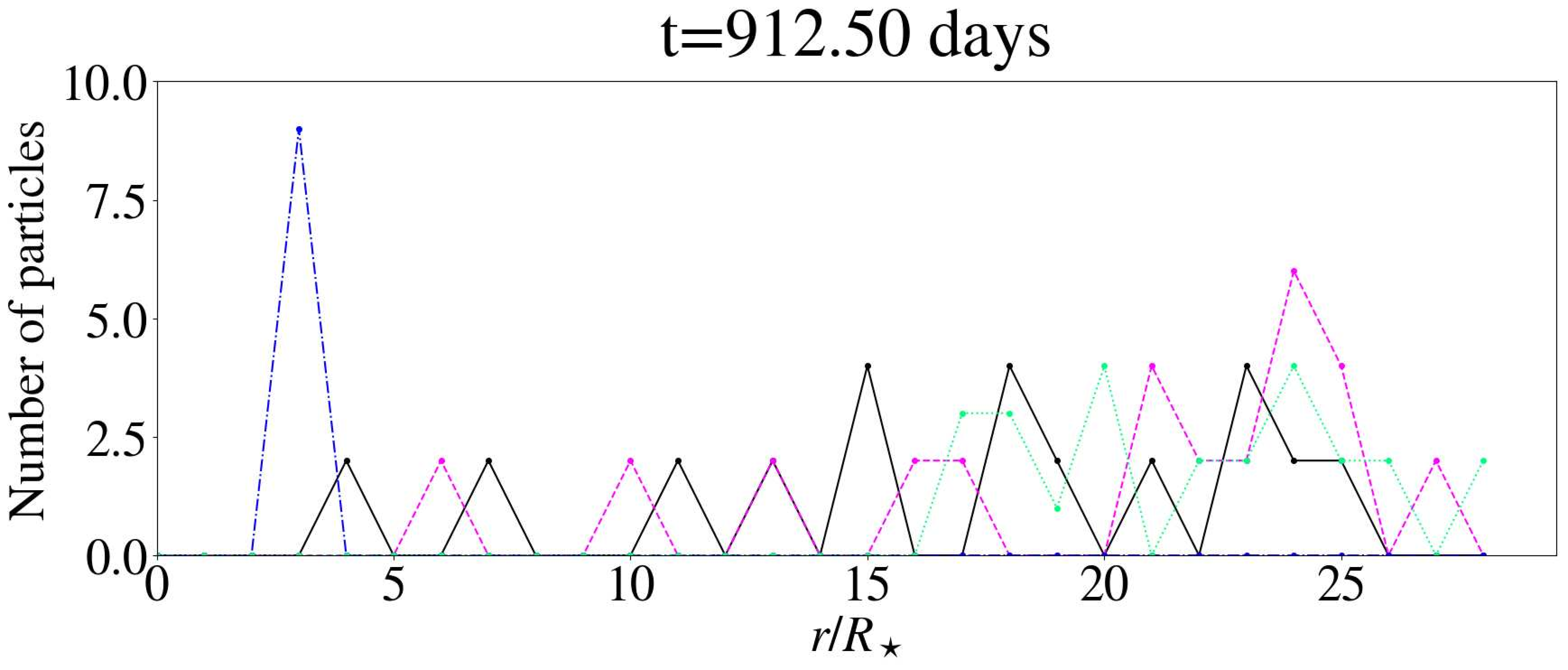}
  \end{minipage}
  \quad
  \begin{minipage}{0.48\textwidth}
    \includegraphics[width = 6cm]{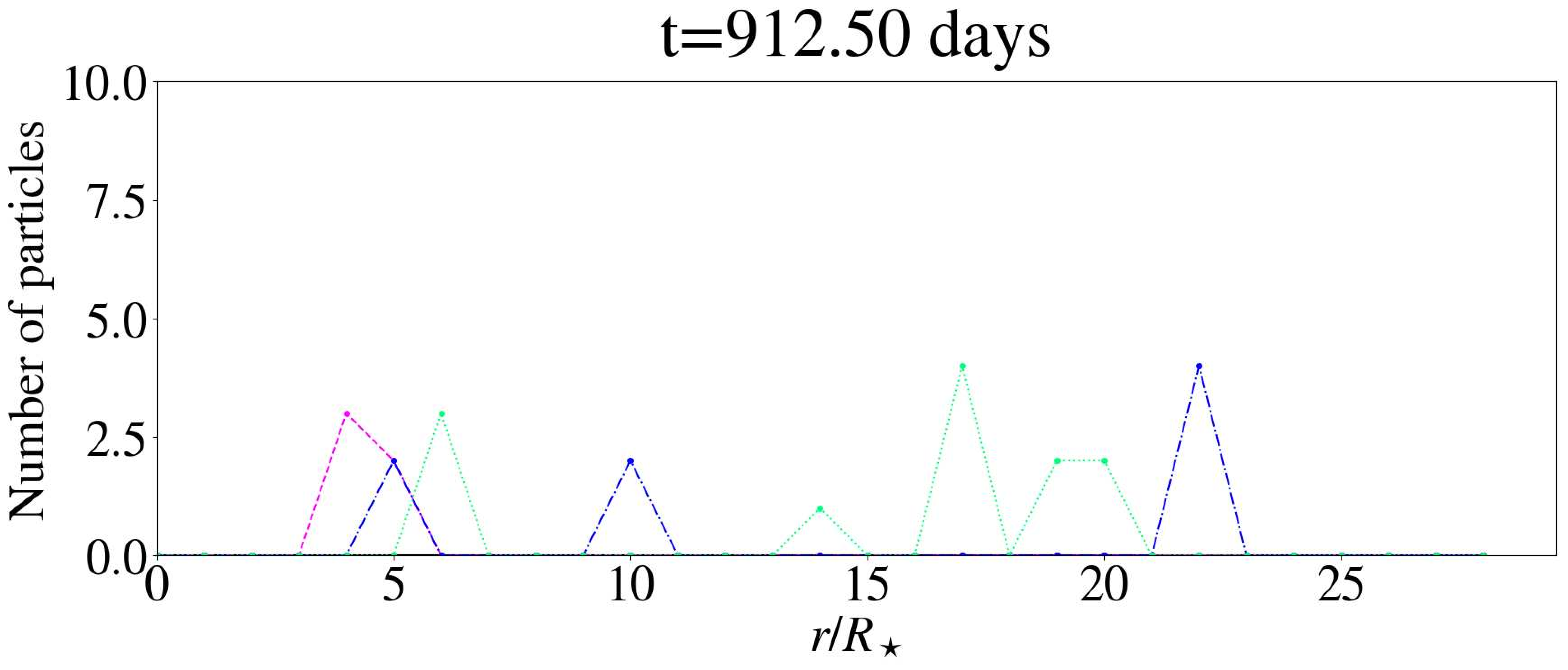}
  \end{minipage}
    \caption{\small{The number of particles at different distances from the
star.  In the left panel is shown a case without backflow and in the right
bottom panel a case with backflow.  Colors correspond to the particles in
Fig 1, with white particles shown in black color line.}} 
\end{figure}

\begin{figure}
    \centering \includegraphics[width = 11cm]{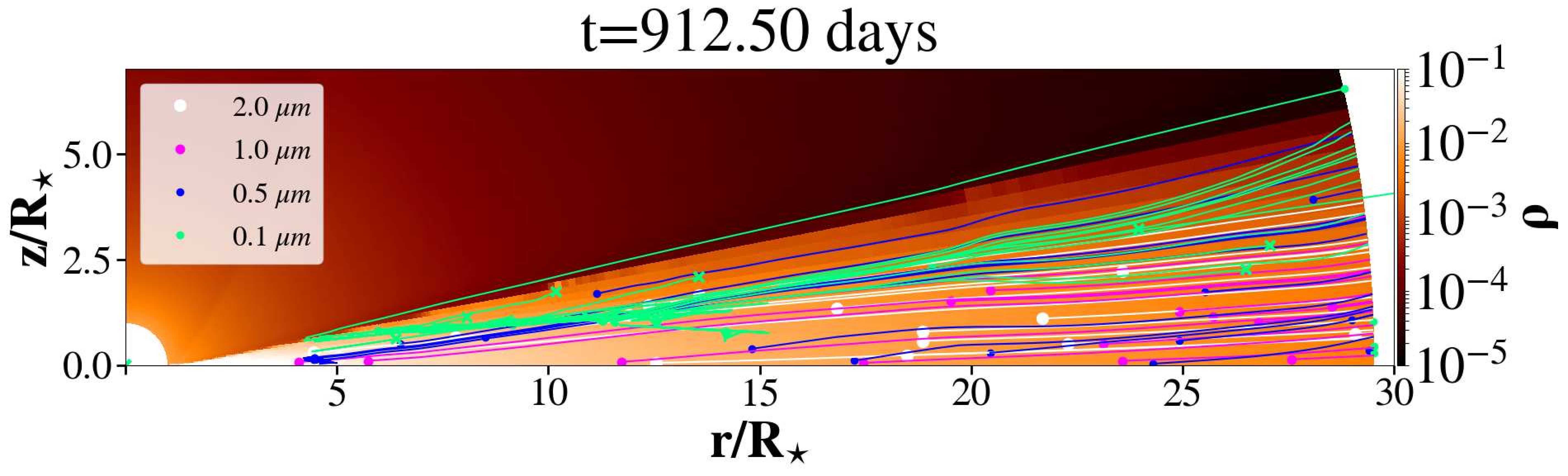}
    \caption{\small{A case with backflow in the disk, but without radiation
    pressure.}} \label{fig:my_label2}
\end{figure}\\
\newpage
\section{Conclusions}

Backflow influences paths of dust particles inside the disk.  The drag force
in the disk backflow and the radiation pressure push particles to the outer
edge of the disk.  In the case without backflow in the disk, particles move
towards the disk equatorial plane and towards the star.  The radiation
pressure in an optically thin disk is preventing the transport of particles
towards the star.

\section{Acknowledgements} 
Work in NCAC Warsaw is funded by a Polish NCN
grant No.2013/08/A/ST9/00795. M\v{C} developed the setup for star-disk
simulations in CEA, Saclay, under the ANR Toupies grant, and his
collaboration with Croatian project STARDUST through HRZZ grant
IP-2014-09-8656 is also acknowledged.  ASIAA (PL and XL clusters) in Taipei,
Taiwan and NCAC (PSK and CHUCK clusters) in Warsaw, Poland are acknowledged
for access to Linux computer clusters used for the high-performance
computations.  We thank the {\sc pluto} team for the possibility to use the
code.

\bibliographystyle{ptapap}
\bibliography{rjarosPTA2019}

\end{document}